# AN OVERVIEW OF CONTEXT-AWARE VERTICAL HANDOVER SCHEMES IN HETEROGENEOUS NETWORKS

Hanan Alhazmi[1] and Nadine Akkari[2]

[1,2]Department of Computer Sciences, King Abdulaziz University, Jeddah, Saudi Arabia
[1]`halhazmi033@stu.kau.edu.sa` and [2]`nakkari@kau.edu.sa`

## ABSTRACT

*As wireless access technologies grow rapidly, the recent studies have focused on granting mobile users the ability of roaming across different wireless networks in a seamless manner thus offering seamless mobility. The different characteristics of each wireless technology with regards to QoS brought many challenges for provisioning the continuous services (audio/video streaming) in a seamless way. In this paper, we intend to review the existing context-aware methods which offered solutions for service continuity. We looked at the types of context information used in each solution. Through this study, it is clear that context awareness plays a significant role in handover process in order to satisfy users demanding seamless services. Therefore, the goal of this paper is to compare the existing methods grouped as general, IMS based, and WLAN/WiMAX solutions in terms of several criteria, such as interworking architecture, service continuity, and QoS provisioning.*

## 1. INTRODUCTION

The continuing growth in wireless technologies has led the wireless world to be integrated and become a ubiquitous environment which is known as Next Generation Wireless Networks (NGWN). This future network will be able to provide service convergence which means services are available to users independently of their location, access technologies, and type of devices. The integration includes different wireless access networks with different characteristics such as IEEE 802.11(WLAN), IEEE 802.16 (WiMAX), GPRS, UMTS using the Internet Protocol (IP) to offer a various range of high data rate multimedia services to end users. The limited coverage range of WLAN makes it difficult to provide "always-on" connectivity services anywhere and anytime. 3G technology offers universal network access but the access rate is very limited. WiMAX can provide high speed internet access in wide area. Therefore, the solutions of WLAN and WiMAX integrated networks can combine their best features while mediating weakness of both networks to provide a complete wireless scheme for offering high speed Internet access to end users.

The integration of NGWN brought many issues. The most challenging one is providing consistent and continuous seamless services while considering Quality of Service (QoS) requirements during the mobility between two different access networks is known as Vertical HandOver (VHO). The ability to maintain service provisioning and avoid flow interruptions while users are roaming is known as service continuity. However, the heterogeneity of wireless networks make service continuity a complex task due to hard issues, such as bandwidth fluctuations and temporary loss of connectivity. This calls for the need of context awareness which claims the full visibility of all characteristics describing service execution environments and enables management operations to adapt service provisioning to current system conditions [1]. To satisfy user needs and improve his/her experience, different context information should be considered. Examples of such information include; user preferences (e.g. preferred network), application requirements, and network conditions.

IP Multimedia Subsystem (IMS) is a standard defined by 3GPP (3rd Generation Partnership Project) which is a multimedia architectural framework for delivering Internet Protocol (IP) multimedia services [2]. IMS facilitate session continuity by offering establishment and control features across heterogeneous network. Also, it provides an access independent architecture which uses Session Initiation Protocol (SIP) as the main signalling protocol. IMS alone cannot fully support service continuity, it needs to collaborate with other mobility solutions for certain areas such as location tracking and network discovery function [3].

In this paper, we intend to review numerous schemes which tackled the context awareness in different ways, and define some of the advantages and drawbacks of each solution. The solutions are classified into three categories: First, general solutions related to context-aware vertical handovers in heterogeneous wireless networks are discussed in section 2. In section 3, we present the IP Multimedia Subsystem (IMS) based solutions which target service continuity. Section 4 presents different integrated WLAN/WiMAX networks solutions. Finally, section 5 presents the conclusion and future work.

## 2. GENERAL CONTEXT-AWARE VERTICAL HANDOVER SCHEMES

Context-aware vertical handover solution in [4] aims to maintain the quality of service of multimedia applications during VHO between WLAN and GPRS networks. Several context parameters are considered such as user profile, network profile, and application requirements that are used in the handover decision. Context information is divided into a static profile and a dynamic profile. The static profile holds the context information that changes rarely while the dynamic profile provides the information about the current user and his/her network. Table 1 shows the dynamic and static profiles used in this solution under "Context Parameters" column. In addition, the solution adopts proxy based architecture for communication stream adaptation and packet loss. It consists of two main components: the context repository and the adaptability manager. The context repository holds the static context information and gathers, manages, and evaluates dynamic context. The adaptability manager makes decisions about the context and determines whether adaptation is needed and a vertical handover must be performed. The handover operation is executed by a protocol between two proxies, one at each network. Moreover, proxies help maintain QoS by double-casting data on both networks and buffering packets. The architecture of this approach is shown in Figure 1. This solution is characterized as a mobile assisted, since useful measurements are gathered from different parts of the system and the terminal. However, this work focuses on context management itself rather than context awareness related to service continuity. Also, it has a drawback of gathering context information at a single point, i.e., the context repository which can form a single point of failure. Moreover, it requires frequent communication between the terminal and the network, rising overhead on the radio link.

In [1], the paper proposes a novel middleware solution called Mobile agent based Ubiquitous multimedia Middleware (MUM), which dynamically personalize service provisioning to the characteristics of the users' environments. MUM adopts context-aware proxy based proactive handover management to transparently avoid service interruptions during both horizontal and vertical handovers (handoffs). This middleware exploits the full visibility of context awareness to guarantee service continuity. In addition, context awareness is categorized as handover awareness, QoS awareness and location awareness, thus providing an original solution for handover prediction, multimedia continuity and proactive readdressing/rebinding. Context

International Journal of Computer Science & Engineering Survey (IJCSES) Vol.2, No.4, November 2011

parameters are shown in Table 1 under "Context Parameters" column. Furthermore, the solution claims that handover management should not further complicate application development also should offer a high flexibility and service specific knowledge usually available only at the application layer. MUM is implemented as testbed, serving Bluetooth and WLAN networks. The architecture of MUM is shown in Figure 2.

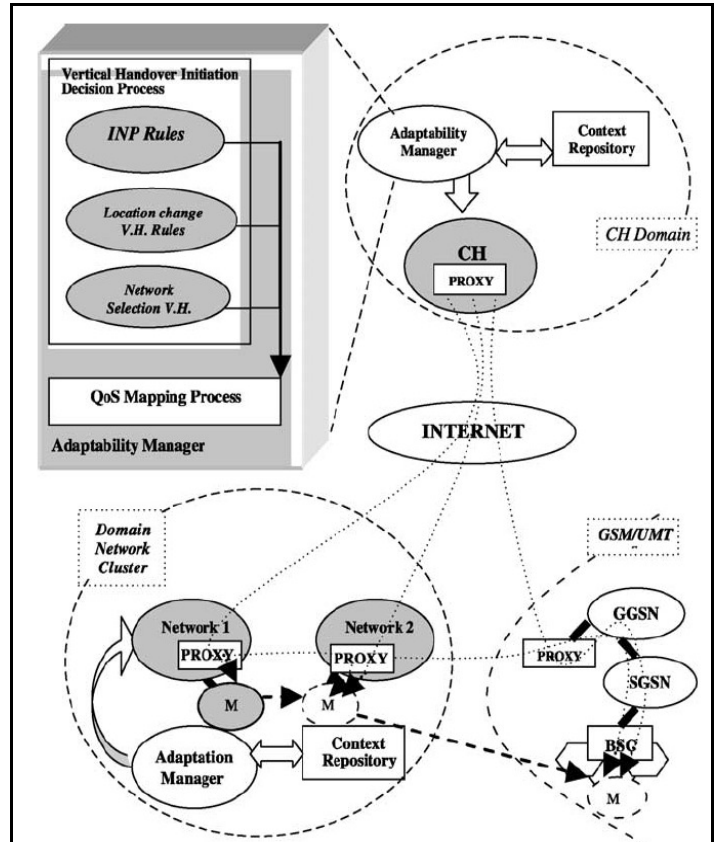

Figure 1.  VHO architecture [4]

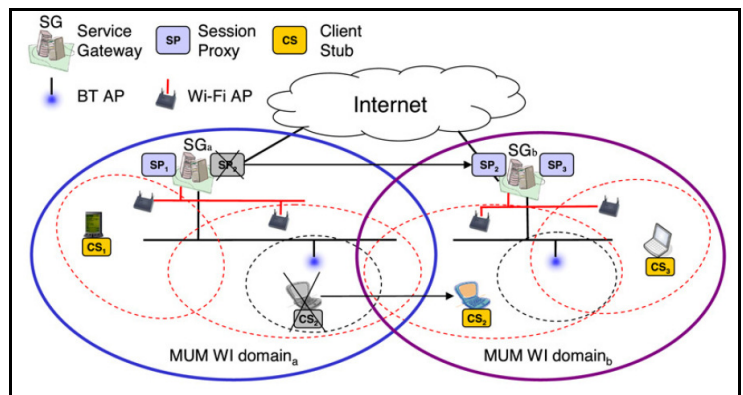

Figure 2.  MUM distributed architecture [1]

The authors in [5] prove that the use of context information helps to improve the delivered application QoS. In their solution, they tackle mobile application connectivity from the context awareness perspective, and describe how high level context information are derived from raw context information. Various context information are taken into account including local host connectivity context, network QoS parameters, application characteristics, as well as user preferences. Furthermore, context information is utilized to develop policies for developing connectivity adaptation. They present a conceptual architecture for policy based adaptive decision of channel selection based on application QoS issues. The combination between different policies and adaptation mechanisms are used to realize adaptive services.

The proposed solution in [6] presents a context-aware profile based middleware to support dynamic service quality management based on the network information in the upper layer, and efficient VHO algorithm with respect to application requirements, user requirements, as well as network information. In this architecture, different modules are designed, such as the application agent which supports the management of service quality as network status changes, and the vertical handover decision manager module. Moreover, this solution defines various profiles related to the context of applications services, user intention, and network status.

In [7], the authors present a context-aware framework in which context information is distributed over many context repositories (a user profile repository, a location information server and a network traffic monitor). It also describes an execution platform for the dynamic deployment and execution of context handling components which minimizes the handover decision time. This purpose is achieved by software agents which are used for the preparation of the collected context data and by the algorithm needed for the handover at the context collection stage. Context information include user's information (location, speed and trajectory) and application QoS. Figure 3 shows the architecture of this solution. This study proves through evaluation that handovers are more efficient when context information is considered. Furthermore, it seems to be flexible for the possibility of using different protocols in exchanging different types of context information and to use different context-aware decision algorithms on mobile terminals.

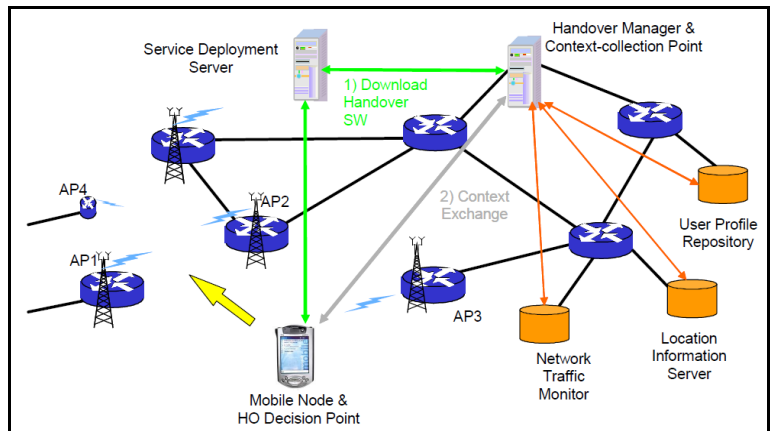

Figure 3. Architecture for context-aware handover using active networking technology [7]

Ahmed et al. have developed and analysed an intelligent handover decision algorithm based on Analytic Hierarchy Process (AHP) algorithm including the session transfer, which takes into account context information resided in the mobile terminal and the network [8]. Session transfer managing is performed by a session transfer scheduling algorithm in order to switch applications to the selected network. Table 1 shows the dynamic and static context information that has been considered in their handover decision model.

In [9], the authors present an integrated framework for mobile nodes (MNs) and core networks to provide a seamless multicast mobility service according user contexts (user states and network states). The parameters of available networks status and host application service information are collected periodically. The context information is stored in the context awareness database to enforce the best strategy for horizontal handover in homogeneous

International Journal of Computer Science & Engineering Survey (IJCSES) Vol.2, No.4, November 2011

networks and for vertical handover in heterogeneous networks. The best one of available access points is selected based on user defined policies and network conditions. Finally, the user devices can receive the most suitable multicast connection and the best application service quality according to system environments.

A mobile service platform with a context-aware middleware is presented at [10] which aims to support vertical handover. It takes as context parameters the user trip information (location and time), device services, network services, user preferences, device specifications, and a given location/time. Once the information is collected, a QoS predictor performs a path prediction to ensure the end to end QoS. The gathered and predicated information is evaluated by AHP algorithm. Due to packet transfer interruptions resulted by handover operation, it is necessary to use buffering techniques to overcome packer losses.

Table 1 shows a comparison of each of the solutions described above in terms of involved networks, service continuity, QoS provisioning and related context parameters.

Table 1. Comparison of general context-aware handover solutions

| Ref. No. | Scope | Involved Networks | Service Continuity | QoS Provisioning | Context Parameters |
|---|---|---|---|---|---|
| [4] | Handover decision, Stream redirection | WLAN - GPRS | ✓ | ✓ | **Static profile:** Device capability, software types, Network interfaces, network QoS and 2D map of network coverage, and user QoS requirements. **Dynamic profile:** Current user location, current QoS network parameters, and Impending Network Profile (INP) |
| [1] | Handover management middleware | Bluetooth - WLAN | ✓ | ✓ | Handover related information (such as type, datalink handover latency and prediction time), Service level Specification (SLS) requirements, user preferences, device characteristics, and network information. |
| [5] | Handover decision, Service adaptation | WLAN - GPRS | ✓ | ✓ | Local host connectivity context, network QoS parameters, application characteristics, and user preferences |
| [6] | Service management middleware, VHO decision | WLAN-GSM/GPRS-Wireless broadband (Wibro) | ✓ | ✓ | **Application profile:** Network requirements for different application service modes. **Working profile:** Information related to the applications (generated by rule-based context-sensitive engine that infers user intention based on the information from various contexts). |
| [7] | Context management, Handover decision | UMTS - WLAN | ✘ | ✓ | User's information (location, speed and trajectory) and application QoS. |
| [8] | Handover decision, Session transfer management | GPRS - WLAN | ✓ | ✓ | **Static:** Device capabilities, service types, QoS requirements of services, user preferences, and provider's profile. **Dynamic:** Running application type, reachable Access points (Aps), and current QoS parameters of Aps. |
| [9] | Handover decision for multicast service | WLAN-3G | ✘ | ✘ | Network status and application service information. |
| [10] | Handover decision | Heterogonous networks | ✘ | ✓ | User trip information, device services, network services, user preferences, device specifications, and mobile location at certain time. |

## 3. IMS BASED SCHEMES

In [11], proactive tight coupling framework is presented which called NetCape (Networking Context aware policy environment). NetCape is policy based context-aware, in what a policy engine runs on the user device. The policy engine gathers information from the user, the network operator, as well as, the environment. There are two kinds of information sources: policy information (user, operator) and environmental information. Environment information consists of the application, session, transport, network, and link layer information. The authors evaluate the proposed method by considering a hierarchical 3G/WLAN network and a user of the 3G networks that has started a video conference with one of his friends. During his conversation the NetCape module decides to perform a hard handover and deliver the ongoing session to a WLAN. The NetCape architecture detects the need for a handover successfully, hence achieving the desired session continuity. This solution only focuses on handover prediction and Mobile IP delay optimization and does not tackle communication streaming adaptation. Moreover, tight-coupling approach requires that any communication from the WLAN domain passes through the core cellular network, with increased communication costs. The architecture of NetCape is shown in Figure 4.

In [12], the authors propose a new approach to integrate mobile Stream Control Transmission Protocol (mSCTP) into IMS based networks in order to facilitate session continuity and to provide mechanisms for service control. It is realized by a proxy based on mSCTP that acts as an anchor point for soft vertical handover of multi interface mobile nodes. This solution does not mainly address IMS service continuity and it considers only the case of make before break which raises power consumption.

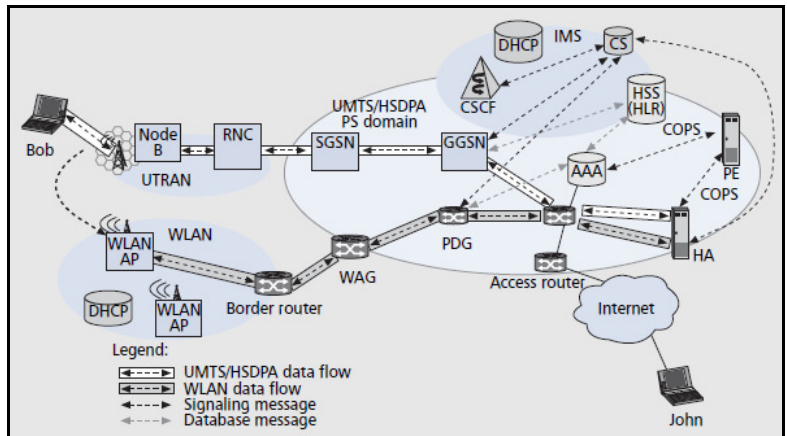

Figure 4. IMS architecture [11]

In [13], they study VHO prediction and session adaptation, as part of a wider architecture called IHMAS (IMS-compliant Handoff Management application server). The solution integrates with existing open source tools for IMS session management and data delivery such as Open IMS Core and Asterisk. The IHMAS architecture (as shown in Figure 5) consists of three main components that perform vertical handover prediction and multimedia session adaptation. Vertical handover prediction is realized by the Vertical Handoff Predictor (VHP) that extends the IMS Client with lightweight and completely decentralized handoff prediction via only local access to client devices. Multimedia session adaptation involves two components: the Application Server for Service Continuity (ASSC), deployed at the MN home network that participates to session signalling to implement vertical handover optimization, and the Adaptation Media Gateway (AMG) which is an application layer media gateway that enables dynamic content adaptation at the mobile node (MN) target network. The solution uses a proactive approach, meaning that it predicts the need for a handover in the client side and the

International Journal of Computer Science & Engineering Survey (IJCSES) Vol.2, No.4, November 2011

necessary operations such session reconfiguration before the initial connection is lost. It is a remarkable that, this solution is the first proposal fully compatible with the IMS and therefore it is ready to be deployed. It adopts an application layer approach for session adaptation. In other words, it exploits application layer media gateways that act as autonomous proxies at client access localities to relieve MN and correspondent node (CN) application components from complex adaptation operations. The results show a significant reduction of the play out interruptions, successful continuous service delivery, and a reduction of the handover time i.e. from 2.230 ms to 220 ms. However, this solution does not mainly addresses IMS service continuity. Besides, it considers only the case of make before break by assuming that there is always enough time for the user equipment (UE) to prepare the connection to the new network before leaving the original one. Moreover, SIP operations (register and invite) need a long time to be achieved.

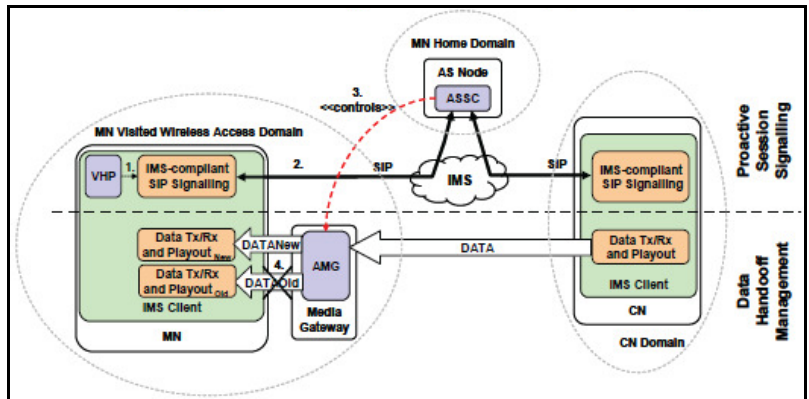

Figure 5. IHMAS architecture [13]

In [14], testbed implementation and evaluation of signalling performance for IMS service continuity using broadband wireless access technologies is performed. Particularly, the authors investigate the signalling aspects of the service in a heterogeneous wireless networks environment in different scenarios. A prototype service continuity application server is developed and used with the UCT IMS Client and the FOKUS Open IMS Core to determine the effect of service continuity signalling on end user experience. The experiments have been implemented on WiMax, WLAN and Fast Ethernet access networks. This work dose not considered context awareness. The testbed is shown in Figure 6.

Table 2 shows a comparison of each of the solutions described above in terms of involved networks, description of each solution, service continuity and context awareness provisioning.

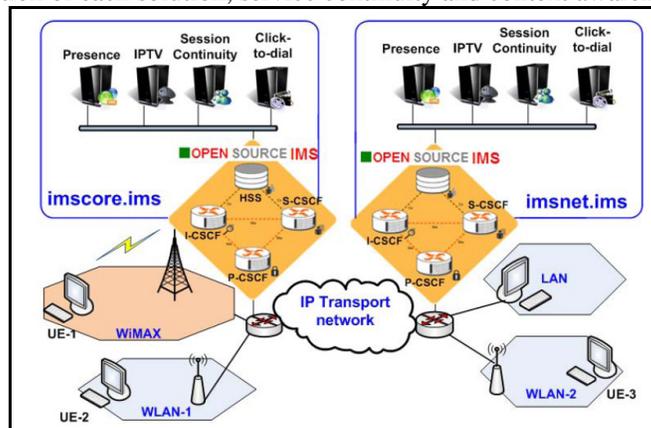

Figure 6. Testbed architecture [14]

Table 2. Comparison of IMS based solutions

| Ref. No. | Description | Involved Networks | Service Continuity | Context Awareness Provisioning |
|---|---|---|---|---|
| [11] | Handover prediction/ handover decision | 3G -WLAN | ✓ | ✓ |
| [12] | mSCTP-based proxy | WLAN-UMTS | ✗ | ✓ |
| [13] | Handover prediction/ session adaptation | WLAN-Bluetooth | ✗ | ✓ |
| [14] | IMS session handovers | WiMax- WLAN- Fast Ethernet | ✓ | ✗ |

## *4. WLAN/WIMAX INTEGRATED NETWORKS SOLUTIONS*

In [15], the paper presents loose coupling integration architecture of WLAN/WiMAX access networks. It also describes several interworking scenarios, where WLAN users with ongoing voice, video and data sessions handover to WMAN (WiMAX), study QoS and performance issues and analyse feasibility of seamless session continuity. In addition, the authors propose a new QoS mapping solution related to the continuity of service between WLAN/WMAN networks. Nevertheless, in this solution, the context awareness is not mainly addressed. The architecture of this solution is shown in Figure 7. The main characteristic of this architecture is to assume two overlapped cells of a WMAN and a WLAN hotspot, where both cells are served by a base station (BS) and an Access point (AP) consecutively. The WLAN cell will extend the size of the WiMax cell. More than one AP could be located at the fringe of the WMAN coverage area.

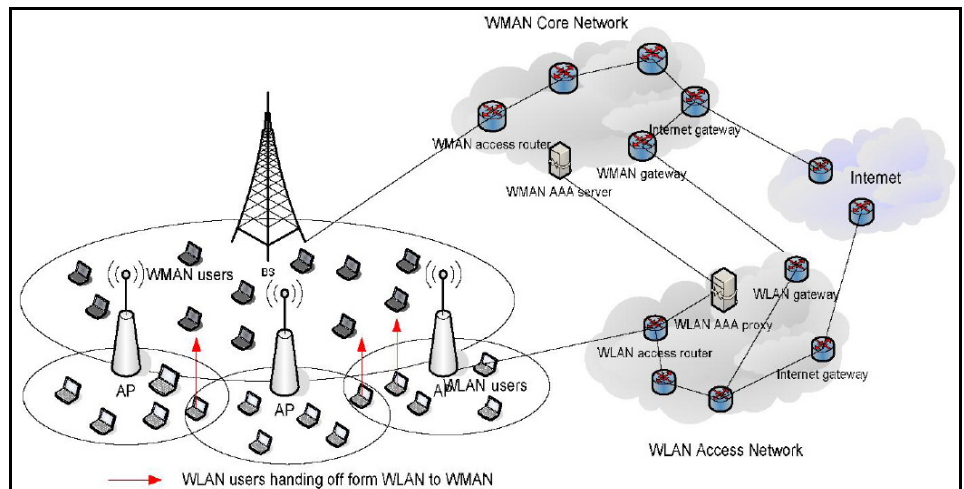

Figure 7. Interworking architecture [15]

Widyo et al have proposed an integration of the terminal management framework with the link layer QoS mechanism to better support various applications over the wireless links [16]. The paper also discusses the scenarios of terminal controlled vertical handover for both the WLAN-WMAN and the WMAN-WLAN handover which use L2 parameters to aid the interface selection and handover decision with considering user preferences. The paper exploits the media Independent Handover (MIH) capabilities to facilitate cross layer information exchange

International Journal of Computer Science & Engineering Survey (IJCSES) Vol.2, No.4, November 2011

between link layer and upper layers. The solution also addresses the QoS mapping performed during L2 mobility. However, this work does not tackle the issue of service continuity.

In [17], the paper proposes a proactive QoS based VHO scheme for integrated WLANs and WiMAX networks. In this work, the service continuity is not addressed. However, the paper aims to present a general VHO management (VHOM) scheme to make VHO decisions which has been designed for each station to monitor the quality of connections and proactively detect network conditions. VHOM selects the network which can provide better service to serve the station rather than a preferred network. They also propose two bandwidth estimation algorithms to predict the available bandwidth in WLAN and WiMAX networks in order to obtain QoS parameters required for handover decision. The main contribution is that a handover process will not only be triggered by unaccepted signal strength but also by unsatisfied QoS parameters. The interworking architecture is shown in Figure 8.

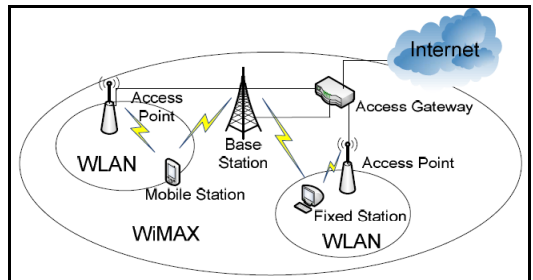

Figure 8. Interworking architecture [17]

In [18], authors introduce a handover decision algorithm by combining MIH QoS model with Multiple Attribute Decision Making (MADM) mechanism to provide seamless mobility between WLAN and WiMAX networks. The proposed model defines QoS parameters mapping among heterogeneous networks, and the service primitives defined in MIH constitute a seamless handover process. The results of the implementation show smaller handover times and lower dropping rate than basic vertical handover methods. However, their work does not guarantee the continuity of the service.

In [19], authors present a vertical handover mobile controlled algorithm for the integration of WLAN, UMTS and WiMAX access networks. The proposed approach is based on user's profile, network's context from MIH and a scoring mechanism which assigns priorities to the QoS parameters (RSS, Bandwidth, delay, etc.) according to the preference of the user. The simulation scenario only involved WLAN and UMTS handover. Through the evaluation, this solution improves QoS parameters including increased throughput and reduced packet loss. However, it does not guarantee service continuity. The architecture of this work is shown in Figure 9.

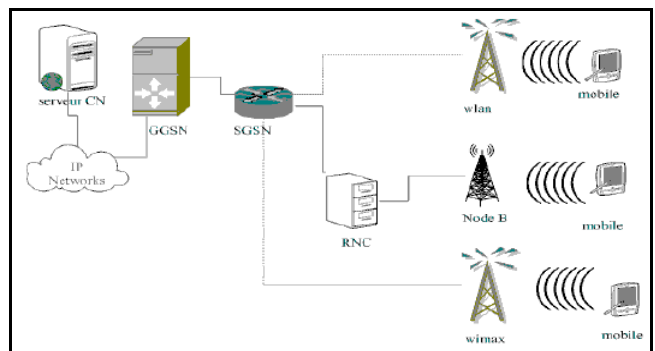

Figure 9. Interworking architecture [19]

Table 3 shows a comparison of each of the solutions described above in terms of interworking architecture, service continuity and QoS provisioning.

Table 3. Comparison of WiMAX/WLAN solutions

| Ref. No. | Scope | Interworking Architecture | QoS Provisioning | Service Continuity | Context Awareness (Parameters/ Methods) |
|---|---|---|---|---|---|
| [15] | Integration architecture | Loose coupling | QoS mapping | ✓ | Burst profile (physical layer characteristics). QoS requirements through DSA-REQ message |
| [16] | Integration architecture/ terminal management framework with link layer QoS | Loose coupling | QoS mapping | ✗ | Link layer parameters:<br>• Channel busy time ratio<br>• Bandwidth request (BW-REQ) loss ratio |
| [17] | Integration architecture/ handover management/ handover decision | Tight coupling | HO decision | ✗ | Service type, available bandwidth, power consumption, cost, throughput and delay |
| [18] | MIH based handover decision algorithm | n/a | QoS mapping | ✗ | **Dynamic QoS information:** Price, bandwidth, delay, PER(packet error rate),jitter |
| [19] | MIH handover decision approach | Tight coupling WLAN-UMTS-WiMAX | HO decision | ✗ | User's profile, network's context from Media Independent Information Service |

## 5. CONCLUSIONS

Next Generation Wireless Networks (NGWN) is a ubiquitous environment that integrates various wireless technologies such as WLAN, 3G, and WiMAX based on IP. One of the most challenging issues of NGWN is providing seamless service continuity while user roaming among different wireless access networks. Context awareness has showed a significant effect to guarantee services continuity. In this paper, we have reviewed numerous context-aware solutions existing in literature. We have classified these solutions into general, IMS based and WLAN/WiMAX integrated networks solutions. A brief description, advantages and drawbacks of each solution were discussed. Moreover, we present a comparative study of the methods and different interworking architectures adopted for service continuity, QoS provisioning and context parameters. Our future work will consist of analysing the optimal solution that will provide service continuity in the context of integrated next generation networks.

### ACKNOWLEDGMENTS

This paper contains the results and findings of a research project that is funded by King Abdulaziz City for Science and Technology (KACST) Grant No. A-S-11-0439

**Authors**

**Hanan Alhazmi,** is a master student at the Department of Computer Sciences, King Abdulaziz University, Jeddah, Saudi Arabia, she received her BSc degree in Computers Sciences (with first honours) from Umm Al-Qura University, Saudi Arabia, July 2007. In Nov 2007, she became a teaching assistant at Mathematics and Sciences department, College of Education (Female), Umm Al-Qura University. She is currently working on her master thesis and expected to graduate at 2012. Her research interests include computer networks, simulations, Next Generation Networks, and WLAN/WiMAX integrated networks.

**Nadine Akkari,** over 8 years of experience in the fields of network engineering, with focus on new emerging technology and all IP network Design in terms of Quality of Services, mobility management and seamless handovers. She studied Computer Engineering at University of Balamand, Lebanon, 1999. She got her Masters in telecommunications networks from Saint-Josepf University, Lebanon in 2001and her PhD in Mobility and QoS Management in Next Generation Networks in 2006 from National Superior School of telecommunications (ENST), Paris, France. Currently she is assistant professor in the faculty of Computing and Information Technology in King Abdulaziz University, Jeddah, Saudi Arabia.